# Multiphoton fluorescence excitation with real intermediary states

**J. David Wong-Campos,**[1,*] **Mackenzie Dion**[1]

[1] *Triplet Imaging, 325 Vassar St., Cambridge, Massachusetts, 02139*
*\*jwongcampos@gmail.com*

**We demonstrate fluorescence generation through a sequential multiphoton process with real intermediary states. Our findings suggest new directions for optical control of previously unexplored molecular excitation pathways.**

Multiphoton excitation has revolutionized microscopy, materials processing, and spectroscopy through precise three-dimensional control at microscopic scales [1, 2, 3] The process relies on the simultaneous absorption of multiple photons via virtual intermediate states, requiring high peak powers, typically from femtosecond lasers, and point scanning systems, limiting throughput and accessibility. Reducing these power requirements while maintaining nonlinear excitation would enable new approaches to study dynamic processes across large volumes at high throughputs. Inspired by optical control of forbidden state transitions in atomic systems [4], we demonstrate an alternative approach using forbidden transitions to triplet states [Fig. 1(a)], enabling red-shifted excitation with both pulsed and continuous-wave sources at lower powers. Using the green-emitting dye Erythrosine B (ErB) in a rigid matrix, we demonstrate this approach by directly populating triplet states [5].

The experimental system is shown in Fig. 1(b). A supercontinuum laser (SC, FIU-15, NKT) provides picosecond pulses at variable repetition rate and wavelength. A dichroic filter (D1, Di02-R594-25x36, AVR) splits the beam into visible (400-600 nm, dashed line) and red/IR (solid line) paths. The visible path incorporates an AOTF (48062-1-.55-1W, Neos) for amplitude control and spectral filtering (F1, FGS900-A, Thorlabs). The red/IR path uses three filters to eliminate spurious visible light (long-pass F2: FGL630M, short-pass F3: FESH1000, Thorlabs; tunable filter pair TFP: LF104550/LF104555, Delta), with amplitude control via a liquid crystal modulator (Mod, LCC1622, Thorlabs) and 6X expansion (Ex1, BE06R, Thorlabs). The beams are combined by dichroic filters (D2: FF495-Di, AVR; D3: T610lpxr-t3-Di, Chroma) and directed to the objective (UPLXAPO 40X, 0.95 NA, Olympus). We detect fluorescence through an emission filter (F4, FF01-572/28-25, AVR) and phosphorescence using a dichroic (D4, FF660-Di-lpxr, Chroma) with a bandpass filter (F5, FF01-679/41-25, AVR). Both signals are focused by tube lenses to a photon counter (FastGatedSPAD, MPD) and digitized (Time Tagger Ultra, Swabian) for lifetime measurements. We characterized excitation at the sample plane using power scans [Fig. 2(b-d)]. Average power ranges were 0-30 μW (blue), 0-20 mW (red, picosecond pulses), 0-50 mW (red, continuous wave), measured at the sample plane.

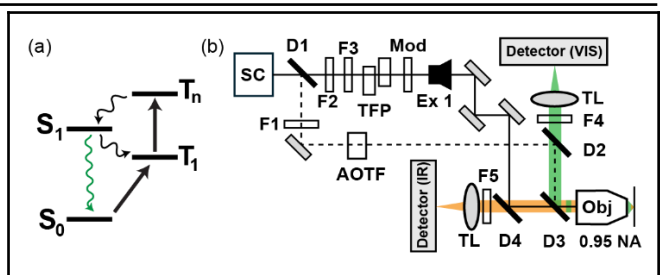

Fig. 1. (a) Proposed energy level diagram of the excitation pathway. (b) Experimental system for broadband excitation and detection of fluorescence and phosphorescence.

The dye sample was embedded in 10% (w/w) polyvinyl alcohol (PVA) in deionized water at 10 μM concentration. A thin film was deposited on a #0 coverslip and dried for 2 hours. This low concentration was chosen to avoid aggregation effects, enabling characterization of the monomeric species [6].

The emission spectra in Fig. 2(a) shows the detection band FL (green-shaded region) and phosphorescence band PH (red-shaded region), which consists of a main fluorescence peak at 565 nm and secondary peak at 680 nm, associated with phosphorescence emission respectively [7]. One-photon excitation at 472 ± 8 nm exhibited typical saturation behavior and a fluorescence lifetime of 724 ± 8 ps [Fig. 2(b-c)] when detecting in the fluorescence detection band (FL). To our surprise, we detected a power-dependent fluorescence signal with single-wavelength excitation at 626 ± 20 nm [Fig. 2(d)]. The measured lifetime decay of 658 ± 8 ps for red excitation was similar to conventional one-photon excitation, confirming emission from a fluorescent state S1 [Fig. 2(e)]. We did not observe saturation effects with red wavelengths. Moreover, red CW laser light at 660 nm (LP660-SF50, Thorlabs) also produced fluorescence [Fig. 2(d)] at four times higher power than the picosecond laser. Similarly to how blue-shifted excitation leads to red-shifted (green) fluorescence through transitions within the singlet manifold, exhibiting the characteristic Stokes shift, the proximity of the 626 nm light to the longer-wavelength phosphorescence peak suggests stimulation of the triplet manifold.

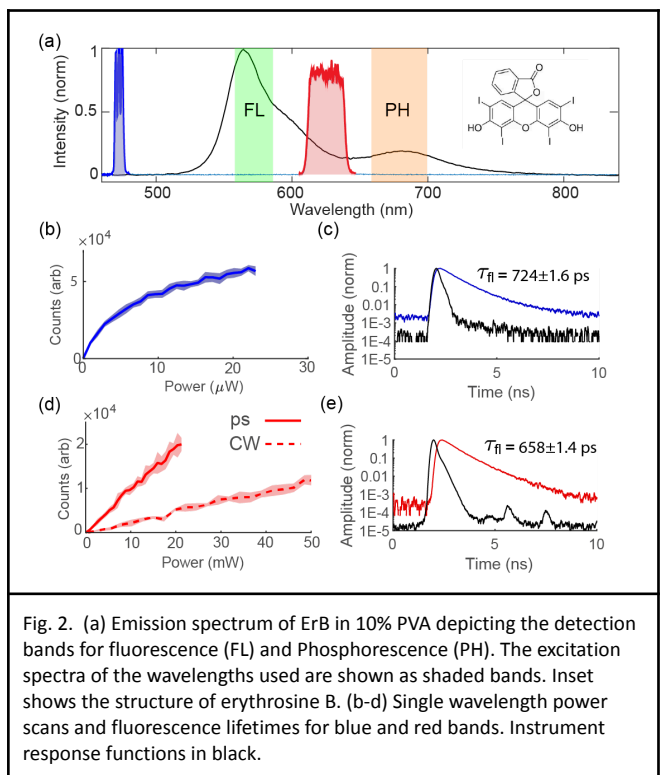

Fig. 2. (a) Emission spectrum of ErB in 10% PVA depicting the detection bands for fluorescence (FL) and Phosphorescence (PH). The excitation spectra of the wavelengths used are shown as shaded bands. Inset shows the structure of erythrosine B. (b-d) Single wavelength power scans and fluorescence lifetimes for blue and red bands. Instrument response functions in black.

We characterized the triplet state through its phosphorescence lifetime [Fig. 3(a)]. When exciting with blue pulses and detecting at the phosphorescence band [PH of figure 2(a)], we observed a fast fluorescence decay plus a constant offset, indicating population buildup in T1 through intersystem crossing from S1. Varying the pulse spacing revealed the offset's decay (Fig 3b), yielding a triplet lifetime of 2.25 ± 0.1 ns. This lifetime, while longer than fluorescence, indicates efficient spin-orbit coupling due to ErB's iodine atoms [8], and confirms the observation of triplet dynamics rather than different molecular species [9].

Next, we investigated the triplet state dynamics using synchronized blue and red pulses at 40 MHz repetition rate [10]. The observed decrease in phosphorescence offset with increasing red power [Fig. 3(c-d)] revealed a depletion saturation power of 0.4 ± 0.1 mW. This behavior demonstrates a sequential excitation pathway: Trapped population in T1 due to the blue light is promoted to Tn with red light, which then undergoes reverse intersystem crossing to return to the fluorescent state S1.

Finally, when exciting ErB with red light alone, the results in Fig. 3(e-f) show no phosphorescence offset. This is because any population reaching $T_1$ through direct $S_0 \rightarrow T_1$ excitation is immediately depleted via $T_1 \rightarrow T_n$. The forbidden nature of the $S_0 \rightarrow T_1$ transition makes it rate-limiting, explaining the linear power dependence before saturation observed in Fig. 2(c). These observations provide strong evidence for direct optical excitation of triplet states.

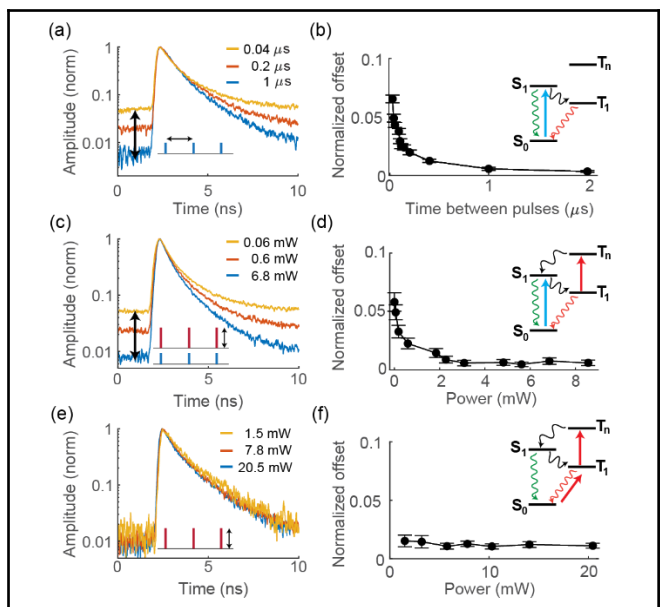

Fig. 3. (a-b) In the phosphorescence detection band (PH), pulsed blue excitation leads to a constant offset, suggesting a build-up of steady-state population in long-lived states. (c-d) At a fixed repetition rate, a combination of blue and red pulses can also decrease the phosphorescence offset. The red light depopulates the long-lived state by excitation to a higher state, which decays to S1 through reverse intersystem. (e-f) Red excitation alone does not produce an offset due to fast depumping from the phosphorescent state T1.

The relaxed temporal requirements of this excitation scheme provide an alternative to multiphoton excitation of dyes commonly used in biology research, where CW two-photon excitation [8] has been used at higher powers (>100 mW). This is evidenced by the modest difference between picosecond and CW excitation, which requires an increase of only 4× more average power in CW mode, highlighting the efficiency of transitions through real intermediate states and challenging the dependence on high peak powers for multiphoton excitation using virtual states. The highly efficient Tn → S1 transition in ErB leads to early saturation of the T1 → Tn step, resulting in a process rate limited by the forbidden S0 → T1 transition. This suggests that chromophores with more balanced transition rates may show stronger nonlinear behavior, motivating future screenings and exploration across different molecular families. Our results showing multiphoton excitation through intermediate long-lived states may enable new approaches in microscopy, lithography, and spectroscopy.

**Funding.** This research was funded by Triplet Imaging, Inc.

**Acknowledgment.** Dalia P. Ornelas-Huerta

**Disclosures.** Authors are cofounders of Triplet Imaging, Inc.

**Data Availability.** Data may be obtained from the authors upon reasonable request